\documentclass[prb,twocolumn,letterpaper,showpacs,amsmath]{revtex4}
\usepackage{graphicx}
\usepackage{color}

\begin{document}
\title{Finite bias charge detection in a quantum dot}
\author{R. Schleser}
\author{E. Ruh}
\author{T. Ihn}
\author{K. Ensslin}
\affiliation{Solid State Physics Laboratory, ETH Z\"urich, 8093 Z\"urich, Switzerland}
\author{D. C. Driscoll}
\author{A. C. Gossard}
\affiliation{Materials Department, University of California, Santa Barbara California 93106}
\date{\today}
\begin{abstract}
We present finite bias measurements on a quantum dot coupled capacitively to a quantum point contact used as a charge detector. The transconductance signal measured in the quantum point contact at finite dot bias shows structure which allows us to determine the time-averaged charge on the dot in the non-blockaded regime and to estimate the coupling of the dot to the leads.

\end{abstract}
\maketitle

\newcommand{\nf}{\normalfont}
\newcommand{\cm}{\,\text{cm}}
\newcommand{\nm}{\,\text{nm}}
\newcommand{\muV}{\,\mu\text{V}}
\newcommand{\mV}{\,\text{mV}}
\newcommand{\V}{\,\text{V}}
\newcommand{\meV}{\,\text{meV}}
\newcommand{\mueV}{\,\mu\text{eV}}
\newcommand{\eVV}{\,\text{eV}/\text{V}}
\newcommand{\Hz}{\,\text{Hz}}
\newcommand{\K}{\,\text{K}}
\newcommand{\mK}{\,\text{mK}}
\newcommand{\T}{\,\text{T}}
\newcommand{\kB}{k_\text{B}}

\renewcommand{\topfraction}{0.9}
\renewcommand{\bottomfraction}{0.9}
\renewcommand{\dbltopfraction}{0.9}
\setcounter{totalnumber}{5}
\setcounter{topnumber}{3}
\setcounter{bottomnumber}{3}
\setcounter{dbltopnumber}{3}

\section{Introduction}
The charge state
of a quantum dot can be read out using a nearby quantum point contact (QPC) as a detector.\cite{Field1993}
Real-time readout has been recently demonstrated
using radiofrequency single electron transistors \cite{Wei-Lu2003} or QPCs.\cite{Schleser2004b,Elzerman2004b}
Together with a spin-charge conversion mechanism,
this is being considered a candidate for a qubit readout scheme \cite{Elzerman2004b} in a future quantum computing device
based on coupled quantum dots.\cite{Loss1998}

A valuable piece of information about a quantum dot's characteristics is the knowledge
about its coupling to each of the reservoirs. This coupling is determined by the electrostatic barrier
forming the constriction and the wave function overlap leading to tunneling.
The latter may strongly depend on the quantum state under consideration
in the dot, which means that the quantum mechanical tunnel coupling has to be determined
for each state individually.
In the Coulomb blockade regime, for the case of single-level transport,
a fit of a transport peak in the Coulomb blockade regime enables one to extract
an effective tunnel coupling
$\Gamma_\text{eff}\equiv\Gamma_\text{S}\Gamma_\text{D}/(\Gamma_\text{S}+\Gamma_\text{D})$
where $\Gamma_\text{S}$ and $\Gamma_\text{D}$ are the couplings to the source and
the drain reservoir, respectively.
However, the ratio $\Gamma_\text{S}/\Gamma_\text{D}$ remains unknown.

In earlier work on finite bias transport \cite{Bonet2002},
e.g. in carbon nanotubes,\cite{Cobden1998}
it was shown that in principle, the individual coupling to both leads of the
dot can be extracted from the current amplitudes at positive and negative bias.
Since these considerations make use of a spin blockade effect, however, they apply only
to the case of spin degenerate states in the dot, and require an even number of electrons
on the dot. In addition, they rely on the absence of cotunneling.
Recently, a method was presented \cite{Leturcq2004} to measure the coupling of a dot
to its reservoirs for a three-terminal quantum dot.

In the following, we present finite bias measurements of transport through a quantum dot,
complemented by simultaneous measurements of the conductance of a nearby,
electrostatically coupled QPC used as a charge detector.
While the latter allows us to determine the time-averaged charge on the quantum dot
even in the non-blockaded regime, the combination of both methods makes it possible to
provide qualitative estimates for
the quantum mechanical coupling of the dot's energy levels to each of the two reservoirs.

\section{Experimental setup}

\begin{figure}[tb]
\includegraphics[width=3.075in]{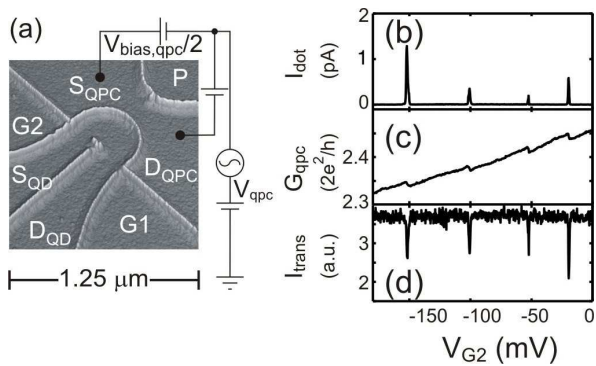}
\caption{(a) AFM micrograph of the structure with designations of 
gates: source (S) and drain (D) of the quantum dot (QD) and the 
quantum point contact (QPC) used as a charge detector; lateral gates 
G1 and G2 to control the coupling of the dots to the reservoirs; 
plunger gate (P) to tune the QPC detector. Only the part of the circuit
related to the readout functionality is included. $V_\text{qpc}$
consists of a dc and a small ac component ($V_\text{qpc,ac}\leq 5\muV$).
(b) Example measurement of the current through the dot. For this
low bias measurements, the voltage across the dot was
$V_\text{\textnormal bias, dot}=10\muV$.
(c) Simultaneous measurement of the conductance through the QPC, 
where each step corresponds to a change of the dot's charge by one 
electron.
(d) Simultaneous measurement of the transconductance $dI_\text{qpc}/dV_\text{qpc}$.
\label{fig1}}
\end{figure}

The sample (see Fig. \ref{fig1}(a)) was fabricated using
surface probe lithography \cite{Held1999, Luscher1999,Nemutudi2004}
on a GaAs/Al$_{0.3}$Ga$_{0.7}$As heterostructure, 
containing a two-dimensional electron gas (2DEG) $34\nm$ below the 
surface as well as a backgate (BG) $1400\nm$ below the 2DEG.
The unstructured 2DEG had a mobility of $(3.5\pm 0.5)\cdot 10^{5}\cm^{2}/{\rm Vs}$
and an electron density $4.6\cdot 10^{11}\cm^{-2}$
at a BG voltage $V_{\rm BG}=-0.5\V$ at $T=4.2\K$.

All measurements were performed in a dilution refrigerator
at a base temperature of $80 \mK$.

Negative voltages were applied to the surrounding gates (G1, G2, 
S$_\text{QPC}$, D$_\text{QPC}$, the latter two also containing the 
charge detection circuit; see Fig. \ref{fig1}(a)), and to the BG, to reduce the charge 
on the dot and close its tunnel barriers. A voltage applied to gate P 
was used to tune the detector QPC to a regime where it is sensitive to 
the charge on the dot.
The QD bias voltage was applied symmetrically (with respect to ground)
across the dot between source (S$_\text{QD}$) and drain (D$_\text{QD}$).

Due to the electrostatic coupling of the QPC to the dot, a change 
in the dot's charge leads to a modification in the QPC's confining potential,
resulting in a change of its conductance.\cite{Field1993} The latter was measured by applying a 
dc voltage and measuring the resulting current.
Each additional electron on the dot leads via electrostatic interaction to a shift
of the QPC conductance's dependence on gate voltage:
$I_\text{qpc,N}(V_\text{gate})=I_\text{qpc,N+1}(V_\text{gate}+\Delta V)$,
where $N$ is the number of electrons on the dot.
If we assume that the QPC is tuned to a regime between two conductance plateaux with an approximately
constant derivative $dI_\text{qpc}/dV_\text{gate}$, then we can subtract $I_{\text{qpc,}N=\text{const}}(V_\text{gate})$
as a background and get a signal proportional to the additional charge on the dot:
$Q\propto I_\text{qpc}(V_\text{gate})-I_{\text{qpc,}N=\text{const}}(V_\text{gate})$.
This technique has recently been applied to investigate the charging behaviour of a double
quantum dot.\cite{DiCarlo2004}

In our measurements, in order to avoid the influence of nonmonotonuous
drift in $I_\text{qpc}$, we measured
the transconductance $dI_\text{qpc}/dV_\text{qpc}$
as an additional quantity, using a lock-in setup at a frequency of $f=31\Hz$. The voltage $V_\text{qpc}$
was composed of a constant dc voltage, used to tune both the dot's and the QPC's chemical potential,
and a small ac voltage used to periodically change the
chemical potential inside the dot by a small amount (of the order of $\kB T$).
The bias $V_\text{bias,qpc}$ across the QPC was applied symmetrically
with respect to $V_\text{qpc}$, in order to minimize its influence on the chemical potential inside the dot.
Figure \ref{fig1}(a) illustrates the QPC related part of the circuit diagram.
Figures \ref{fig1}(b)-(d) show the correlations between the three quantities
in a simultaneous measurement:
at the position of a Coulomb blockade peak in the current through the dot, a kink appears
in the QPC's conductance, corresponding to a change in the dot's charge by one.
In the transconductance $dI_\text{qpc}/dV_\text{qpc}$, a dip is observed.
$dI_\text{qpc}/dV_\text{qpc}$, being a derivative of the QPC current, differs from
$dI_\text{qpc}/dV_\text{G2}$ only by a constant background and
a factor given by the ratio of the corresponding lever arms, 
$\alpha_\text{G2}/\alpha_\text{qpc}$. This ratio should depend only weakly on gate voltages.
One can therefore obtain the time-averaged charge on the dot by
integrating the measured value $dI_\text{qpc}/dV_\text{qpc}$ with respect to $V_\text{G2}$,
(after subtracting a constant and a weak linear background to compensate
for the gate voltage dependence of $\alpha_\text{G2}/\alpha_\text{qpc}$) and normalizing the resulting
steps to unity.

\section{Time averaged charge detection}

\begin{figure}[tb]
\includegraphics[width=3.075in]{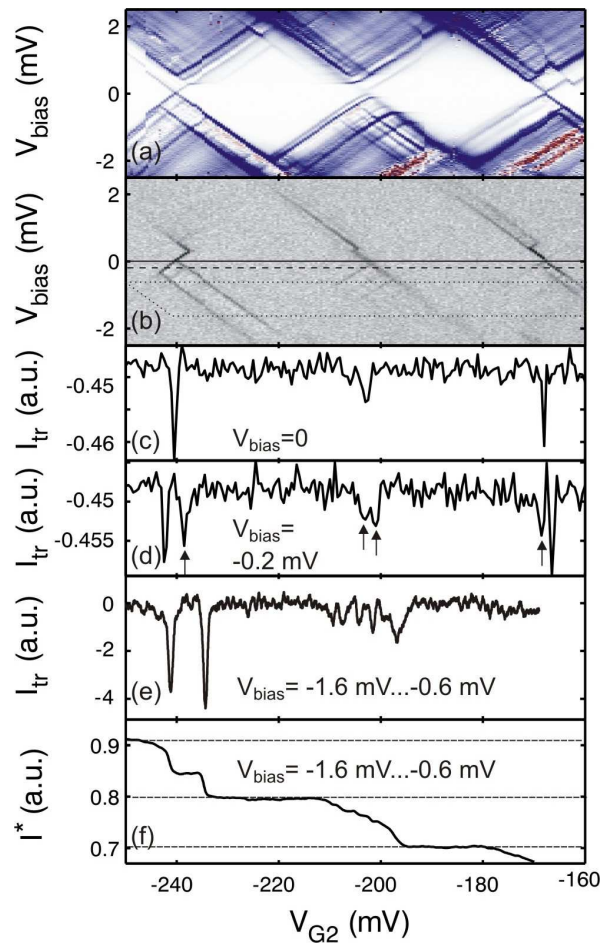}
\caption{(Color online) (a) Finite bias transport measurement through the dot at a magnetic field of
$B=0.1\T$. Dark regions represent larger positive [blue] or negative [red] differential conductance.
(b) Corresponding measurement of the transconductance. Dark regions represent larger
values.
(c) Single transconductance trace for low bias ($V_\text{bias}=20\muV$, see upper horizontal line in (b)).
(d) Single transconductance trace for $V_\text{bias}=-0.2\mV$ (see dashed horizontal line in (b)).
(e) Averaged transconductance.
Averaging was performed over a set of bias voltages $-1.6\mV<V_\text{bias}<-0.6\mV$.
Individual traces were laterally shifted
with respect to each other so that the two lines visible in the lower left part of (b) (parallel to the diamond edges in (a))
yield two distinct sharp peaks. The region delimited by dotted lines in (b) marks the range over which averaging took place.
Furthermore, a constant and a linear background were subtracted.
(f) Averaged transconductance integrated with respect to $V_\text{G2}$.
\label{fig2}}
\end{figure}

In Fig. \ref{fig2}(a), we present finite bias measurements of the dot's
conductance. From these, estimates for the charging energy $E_c\approx 1.6\meV$ 
and the mean level spacing $\Delta\approx 300\mueV$ were
extracted. Fig. \ref{fig2}(b) shows a simultaneous
measurement of the QPC's transconductance.

In the transconductance plot, diagonal lines with the same slope
as the diamond boundaries in the dot conductance plot are observed, marking
a change of the dot's time-averaged charge.
The lines correspond to the alignment of an energy level in the dot with
either the source (negative slope) or the drain (positive slope)
reservoir, while their intensity contains information
about the magnitude of the change in charge.

Figure \ref{fig2}(c) shows the transconductance at low bias.
Dips in the signal at the gate voltages of the Coulomb blockade peaks
correspond to the change in electronic charge by one elementary charge each.
For a finite bias $V_{\rm bias}=-0.2 \mV$ (Fig. \ref{fig2}(d)),
the peaks split, illustrating
that the time-averaged charge on the dot changes in steps smaller
than the unit charge $e$.

These steps in charge can be directly visualized by integrating over
the transconductance (Fig. \ref{fig2}(e)-(f)).
In order to improve the precision of the charge determination, the transconductance
was averaged over a finite bias range prior to numerical integration. A constant
(corresponding to the direct coupling between the in-plane gates used for ac excitation and the QPC)
and a linear term (corresponding to the gate voltage dependence
of $\alpha_\text{G2}/\alpha_\text{qpc}$) were then subtracted. To conserve the steepness
of the steps near $V_\text{G2}\approx -240\mV$, the individual traces were laterally 
shifted with respect to each other before averaging. As a consequence, the step corresponding to 
the second Coulomb peak is broadened, since for this peak the slope of the corresponding
lines in Fig. \ref{fig2}(b) is different.

The large steps marked by dashed horizontal lines in Fig. \ref{fig2}(f), corresponding to
a change in charge by one elementary charge each, have almost identical
height, as one would expect. Charge rearrangements during the measurements, one of which is visible at
$V_{\rm G2}\approx 235\mV$ in Fig. 2(a), might contribute to errors in the numerical integration
and thus may lead to small differences in total step height.

To verify the usefulness of the integrated signal as a measure for the mean charge on the dot,
we compared the step size related to a single electronic charge for several Coulomb peaks
(see Fig. \ref{fig2}(f)) and for different bias ranges between zero and the height of a Coulomb
diamond.
The difference in step height did not exceed 10 percent if a sufficient number of
traces was taken into account to reduce the statistical error. Even in the high-bias regime
$1.2\mV<V_{\rm bias}<1.6\mV$ near the border of the Coulomb-blockaded bias range, 
where the lines in the transconductance seem to gradually disappear,
the integrated transconductance signal gives a good measure of the mean charge,
which means that line broadening compensates for the lower amplitude.

\section{Estimation of dot-lead coupling: non-degenerate case}

From the information about the mean charge in the non-blockaded region, it is possible
to extract the coupling of individual states to each lead separately
by using a rate equation approach:\cite{Beenakker1991}
if a single energy level lies within the (thermally broadened) bias range between the Fermi
level of both leads,
the mean dwell time of the $N$th electron in the dot is determined by the relative value
of the couplings $\Gamma_{{\rm S},N}$ and $\Gamma_{{\rm D},N}$
to both leads. The $\Gamma_{{\rm i},N}$'s account for the height of the tunnel barriers $i$,
the wave function overlap between the dot and lead, and the density of states inside the leads, which we
assume to be constant.
The occupation probability of the energy level $N$ then becomes
\begin{equation}
P(N)=\frac{\Gamma_{{\rm S},N}f_{\rm S}(\mu_N)+\Gamma_{{\rm D},N}f_{\rm D}(\mu_N)}{\Gamma_{{\rm S},N}+\Gamma_{{\rm D},N}}\;{\rm ,}
\label{eqOccupation}
\end{equation}
where $f$ stands for the energy distribution in the leads (which we assume to be the Fermi distribution) and
the indices S and D stand for source and drain, respectively. For a bias voltage
$\left|V_{\rm bias}\right|=\left|E_{\rm F,S}-E_{\rm F,D}\right|>\!\!>\kB T$ and a chemical potential $\mu_N$ not inside
the thermally broadened ranges around the Fermi energies of both leads,
we have 
$f_{\rm S}(\mu_N)=1$,
$f_{\rm D}(\mu_N)=0$, and
this expression becomes
$P^{+}(N)=\Gamma_{{\rm S},N}/(\Gamma_{{\rm S},N}+\Gamma_{{\rm D},N})$
for $E_{\rm F,S}>E_{\rm F,D}$ (positive bias) and
$P^{-}(N)=\Gamma_{{\rm D},N}/(\Gamma_{{\rm S},N}+\Gamma_{{\rm D},N})$
for $E_{\rm F,S}<E_{\rm F,D}$ (negative bias).

Since the non-integer (fluctuating) part of the mean charge on the dot
in the level $N$ is given by
$eP^{\pm}$, this connects the integrated transconductance value to
the ratio $\Gamma_{{\rm S},N}/\Gamma_{{\rm D},N}$:
\begin{equation}
\frac{\Gamma_{{\rm S},N}}{\Gamma_{{\rm D},N}}=\frac{P^{+}(N)}{P^{-}(N)}=\frac{P^{+}}{1-P^{+}}\,.
\label{eqGammaRatio}
\end{equation}

Together with the expression for the transport current
\begin{equation}
I=-e\frac{\Gamma_{{\rm S},N} \Gamma_{{\rm D},N}}{\Gamma_{{\rm S},N}+\Gamma_{{\rm D},N}}
\label{eqCurrent}
\end{equation}
it is possible to determine the numerical
values of $\Gamma_{{\rm S},N}$ and $\Gamma_{{\rm D},N}$.

The analysis of the transconductance signal
becomes more involved if more energy levels inside
the quantum dot are contributing to transport.
It is thus preferable to start the analysis in the
regime of single-level transport,
i.e. $V_\text{bias}<(\Delta-k_{\rm B}T)$.
In addition, the maxima in the transconductance signal associated with source and drain
alignment of the level must be clearly separable,
i.e. $V_\text{bias}>k_{\rm B}T$. Those two requirements limit
the bias range over which the integration
of the transconductance peak can be performed on a given data set,
especially when excited states
are present near the ground state.
To apply the method described above, we have to verify that there exists a finite bias range
where these conditions are met.

We analyzed the low finite bias regime around
three conductance peaks in a range where the system was most stable,
for five different magnetic fields applied, to change the shape of the dot's wave
functions and their overlap with the leads. In our analysis, we encounter two qualitatively
different situations in the low bias transconductance regime:
No branching at zero bias (this section): single line with constant slope crossing the zero bias line,
possibly branching at finite bias.
Branching at zero bias (degenerate case, discussed in the next section).

\begin{figure}[tb]
\includegraphics[width=3.075in]{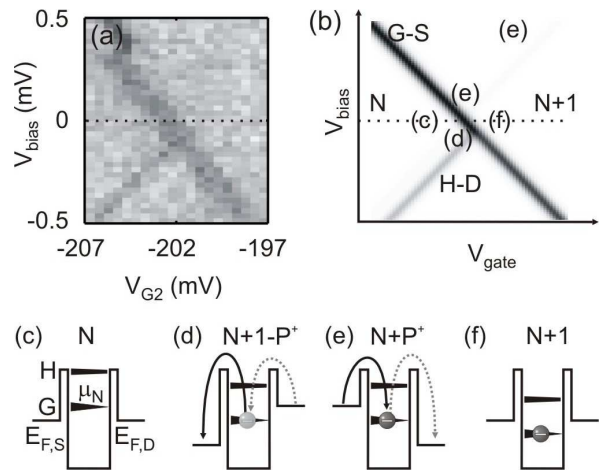}
\caption{
(a) Detail from Fig. \ref{fig2}(b), showing a broadened line with
negative slope at zero bias, and a splitting at finite bias.
(b) Illustration of the connection between the coupling
of a single state (G) to source (S) and drain (D) and the mean
charge on the dot.
This representation was generated using a simulation
based on a rate equation approach,\cite{Beenakker1991}
and is compatible with our experimental observations (with the exception of
line broadening due to strong coupling, which is not included in the theory).
Our model for this case includes two single-particle levels: an asymmetrically coupled
state accounting for the imbalance of line weight observed at positive and low negative bias
(where the line with positive slope, corresponding to
alignment of G with D, is practically absent),
and a higher energy level H with more symmetric coupling, responsible for
the branching at finite negative bias.
We used coupling asymmetries
$\Gamma_{{\rm D,G}}/\Gamma_{{\rm S,G}}=1/50$ for the level G,
and $\Gamma_{{\rm D,H}}/\Gamma_{{\rm S,H}}=1/4$ for the level H.
In additon, we set
$\Gamma_{{\rm D,G}}+\Gamma_{{\rm S,G}}=\Gamma_{{\rm D,H}}+\Gamma_{{\rm S,H}}$.
The diagonal line (G-S) with negative slope
corresponds to the alignment of a single energy level (G) with the source (S) contact,
while the line (H-D) with positive slope in the negative bias range marks an alignment
of an excited state (H) with the drain (D) contact. Letters in brackets (c)-(f)
located in the blockaded (left and right) and non-blockaded (top and bottom)
regions refer to the corresponding figures. These illustrate how
the time-averaged electron number depends on bias and chemical potential of the dot
for the case of an asymmetric (here: $P^{+}>\frac{1}{2}$) coupling.
\label{fig3}}
\end{figure}

An example of the first case is encountered, e.g., at the second peak in Fig. \ref{fig2}(a)/(b)
(see detail in Fig. \ref{fig3}(a)), in the low bias regime:
near $V_{\rm bias}=0$, we observe a broadened line with negative slope
in the transconductance, while a similarly broadened, but comparatively weak maximum
is measured in the transport current (Fig. \ref{fig2}(a)).

Figure \ref{fig3}(b) shows a schematic representation of Fig. \ref{fig3}(a),
illustrating which line corresponds to the alignment of the dot's chemical potential
with source (S) and drain (D). Figures \ref{fig3}(c)-(f) illustrate how coupling and level
alignment determine the mean electron number on the dot. Note that in Fig. \ref{fig3}(e),
an asymmetrically coupled level can become nearly 100 percent
occupied even in the non-blockaded regime.

The observation near $V_{\rm bias}=0$ is compatible with the
presence of a single energy level strongly coupled to the source reservoir (leading to enhanced
broadening, compared to a Coulomb blockade peak width of purely thermal origin), but extremely weakly
coupled to the drain contact. This would explain the observed low zero-bias
transport current and the fact that within our measurement resolution, no line with positive
slope corresponding to an alignment to the drain contact is observed in the transconductance.
At finite negative bias, i.e. once a more symmetrically coupled excited state
becomes accessible (see finite bias region around middle peak in Fig. \ref{fig2}(a)),
transport is strongly enhanced.

Solving equations \ref{eqOccupation} and \ref{eqCurrent} for
$\Gamma_{{\rm S},N}$ and $\Gamma_{{\rm D},N}$ yields
an upper bound for the ratio of the coupling constants $\Gamma_{{\rm D},N}/\Gamma_{{\rm S},N}\leq 1/50$.
From the measured value of the transport current in the finite bias regime
(averaged over positive and negative bias), we obtain an estimate for 
$\Gamma_{{\rm S},N}\Gamma_{{\rm D},N}/(\Gamma_{{\rm S},N}+\Gamma_{{\rm D},N})$, so that both are
uniquely determined within the errors mainly resulting from charge measurement uncertainties. In fact,
for this very asymmetric case, we might assume that level broadening is caused entirely
by the more strongly coupled source lead. The full width at half maximum of the Coulomb peak
is of the order $w_{\rm FWHM}=130\mueV<\Delta$, yielding a value for $\Gamma_{{\rm S},N}$
which suggest an even stronger asymmetry
$\Gamma_{{\rm D},N}/\Gamma_{{\rm S},N}\leq 1/1000$.

At finite bias, a splitting of the line in the transconductance is observed, suggesting that
an excited state of the dot more symmetrically coupled to both leads governs its mean occupation.
Inserting this into a numerical simulation (see below) also yields qualitative agreement with the observed
finite bias dot current, which shows a larger step change at the positive than at the negative bias corresponding to
the excitation energy.

\section{Estimation of dot-lead coupling: degenerate case}

\begin{figure}[tb]
\includegraphics[width=3.075in]{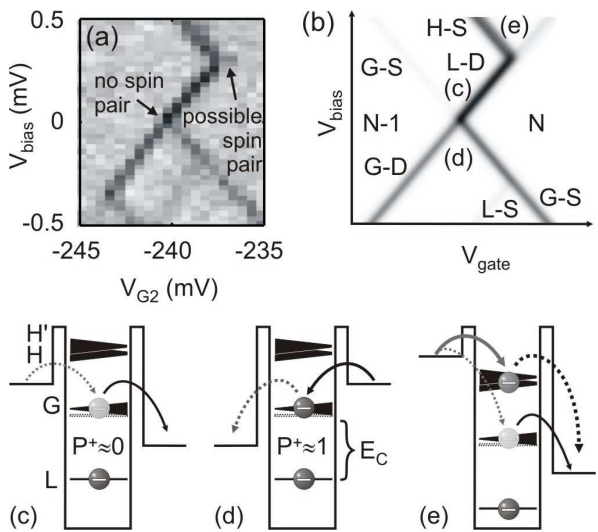}
\caption{(a) Detail from Fig. \ref{fig2}(b), showing discontinuity/splitting of lines
in the transconductance signal at zero bias.
(b) Schematic representation of (a) generated by a numerical simulations.
To explain the observed line splitting, a simplified model spectrum
with four states is used, consisting of two pairs of (quasi-)degenerate levels.
Small letters in brackets refer to Figures (c)-(e), where
the spectrum is depicted for different values of gate and
bias voltage, illustrating how the interplay of several levels with different couplings to the leads
influences the time-averaged occupation of the dot.
The labeled lines in (b) refer to the alignment of a level (L, G, or H/H', see Fig. (c))
with either source (S) or drain (D).
(c) For low positive bias, charge passes through
the degenerate levels G and L. Due to the asymmetric coupling,
the mean charge remains low until alignment with the drain contact.
(d) For small negative bias, the asymmetry in the coupling of levels L and G leads to
a finite mean charge on the dot as soon as they align with
the drain's chemical potential (see line G-D in Fig. \ref{fig4}(b)).
(e) In this parameter range, the mean charge reaches a high level, due
to the asymmetric coupling of the levels H and H' that trap charge entering through the source
contact. 
\label{fig4}}
\end{figure}

In a number of cases the lines
in the transconductance plot branch at zero bias (leftmost peak in Fig. \ref{fig2}(b), see detail
in Fig. \ref{fig4}(a)). This splitting is not compatible with the participation of only one individual single-particle level
in transport.
At finite positive bias, a further splitting occurs and a line with opposite bias dominates,
suggesting predominant coupling towards the source contact.

From our transconductance measurements, the mean occupation was determined 
by integrating over a voltage range in $V_{\rm G2}$ in the transconductance plot, summing
over a range of bias voltages to reduce the statistical errors. Normalization was performed by integrating
over the whole non-blockaded gate voltage range, summing over the same bias range.

To elucidate the origin of the low bias results (depicted schematically in Fig. \ref{fig4}(b)), 
we performed analytical calculations using a rate equation approach 
based on the framework of Beenakker's theory of sequential 
electron tunneling.\cite{Beenakker1991}

We assumed that a model involving at least two quasi-degenerate states might describe our findings,
and therefore extended equations \ref{eqOccupation}-\ref{eqCurrent} to the case of
two degenerate levels, involving  four
coupling values $\Gamma_{i,j}$ ($i=\text{S},\text{D}$ for source and drain; $j=1,2$).

To extract a meaningful result despite the large
relative errors in the charge input values, we express the $\Gamma_{i,j}$ in
terms of a global multiplicative factor $\Gamma_0=\sqrt[4]{\prod_{i,j}\Gamma_{i,j}}$,
the symmetries of each of the two states 
$S_i=\sqrt{\frac{\Gamma_{\text{S},i}}{\Gamma_{\text{D},i}}}$
and the relative weight of the two states 
$V=\sqrt[4]{\frac{\Gamma_{\text{S},1}\Gamma_{\text{D},1}}{\Gamma_{\text{S},2}\Gamma_{\text{D},2}}}$,
expressed as the ratio of the geometrical averages of their couplings. The resulting simple expressions for the $\Gamma_{i,j}$ become:
$\Gamma_{\text{S},1}=\Gamma_0 V S_1$, $\Gamma_{\text{D},1}=\Gamma_0 V / S_1$,
$\Gamma_{\text{S},2}=\Gamma_0 S_2 / V$, $\Gamma_{\text{D},2}=\Gamma_0 / (V S_2)$.

Using the four charge and transport values at low positive and negative bias
as input values, the expressions for current and occupation can be solved for the
four new parameters. The numerical values obtained are
$\hbar\Gamma_0=(2.0\pm 0.7)\mueV$, $V=1.23\pm 0.50$, $S_1=1.10\pm 0.15$, $S_2=0.22\pm 0.12$.
This means that one of the two levels has approximately symmetric coupling, while
the second is predominantly coupled to the drain contact. This state must therefore have a very asymmetric 
distribution of the wave function amplitudes to the two reservoir connections, since
we learn from the slopes observed in Fig. \ref{fig2}(b) that the dot as a whole predominantly couples
to the source reservoir, suggesting an asymmetry in the tunnel barriers themselves.
Since the numbers show that the two states have different symmetry properties,
the most obvious possibility of a spin pair has to be excluded in this case.

The method described above can be extended to determine the coupling properties
of higher energy states by incrementally solving the more intricate rate equations
involving a larger number of levels, using measurements at higher bias and
the previously calculated $\Gamma_{i,j}$ values as an input. The major advantage
of this procedure is that it correlates data measured at a constant value
of the controlling gate, changing only the bias of the dot:
due to spectral scrambling, the information obtained 
in this way is rarely accessible by comparing successive Coulomb peaks.

We perfomed this analysis using numerical calculations based on the same
aforementioned rate equation approach.
Even though the input values include already a certain error,
some qualitative statements can be derived:
the most important one concerns the feature visible in Fig. \ref{fig4}(a),
showing a change in slope in the right positive bias line occuring at finite bias.
The fact that the line with positive slope is discontinued
is not compatible with the presence of a single excited state.
Again, two quasi-degenerate states are necessary to explain the scenario
observed. The coupling of these two states has to be asymmetric with 
predominant coupling to the source contact.
The measurements are compatible
with the two states having identical  coupling values, leaving open
the possibility of a spin pair.
Fig. \ref{fig4}(b) shows a numerical simulation using the
values $\Gamma_{i,j}$ from the analytical calculations above
for the two lower states (responsible for the low bias behaviour).
The remaining $\Gamma_{i,j}$ were determined numerically by
iteratively comparing experiment and simulation.

The resolution of our charge detector is limited:
all the transconductance measurements presented above were performed
in a regime in which the QPC's conductance remained between $2e^2/h$ and $4e^2/h$.
Slight changes in the lithographic pattern for future structures
will permit us to approach the tunneling regime,
resulting in a tenfold increase in charge
sensitivity and a more precise determination of the mean charge and the coupling strengths.

\section{Conclusions}
We have demonstrated how a QPC charge detector for a nearby quantum dot
can be used to extract information about the average charge in the non-blockaded
finite-bias regime.
Together with the conductance measurement through the dot itself,
this allows us to extract qualitative and quantitative
information about the coupling of single energy levels to the
source and drain reservoirs.

Financial support from the Schweizerischer Nationalfonds is gratefully acknowledged.

\newpage
\bibliographystyle{apsrev}

\end{document}